\documentclass[a4paper]{article}
\usepackage{fullpage}
\usepackage{amsmath}
\usepackage{amsfonts}
\usepackage{theorem}

{\theorembodyfont{\rmfamily}
\newtheorem{definition}{Definition}[section]}
{\theorembodyfont{\rmfamily}
}
\newtheorem{corollary}{Corollary}[section]

\usepackage{url}
\def\openbox{\mbox{\textvisiblespace}}
\makeatletter
\newcommand{\card}{\mathop{\operator@font card}}
\makeatother
\begin{document}
\frenchspacing
\title{A (Basis for a) Philosophy of a Theory of Fuzzy Computation}
\author{Apostolos Syropoulos\\ Xanthi, Greece}
\date{}
\maketitle
\begin{abstract}
Vagueness is a linguistic phenomenon as well as a property of physical objects. Fuzzy set 
theory is a mathematical model of vagueness that has been used to define vague models 
of computation. The prominent model of vague computation is the fuzzy Turing machine. This
conceptual computing device gives an idea of what computing under vagueness means,
nevertheless, it is not the most natural model. Based on the properties of this
and other models of vague computing, it is aimed to formulate a basis for a philosophy 
of a theory of fuzzy computation.
\end{abstract}
\section{Introduction}
Following Raymond Turner~\cite{sep-computer-science}, who starts his entry on the 
{\em Philosophy of Computer Science} in the Stanford Encyclopedia of Philosophy by quoting
Stewart Shapiro~\cite[p.~525]{shapiro1983} in order to explain in a nutshell what is
the philosophy of computer science, I will use the same excerpt from Shapiro's paper
to give a general idea of what a philosophy of {\em fuzzy} computation should be. More 
specifically, given an arbitrary field of study $X$, Shapiro notes that 
\begin{quote}
\ldots the main purpose of a given field of study is to contribute to
knowledge, the philosophy of $X$ is, at least in part, a branch of 
epistemology. Its purpose is to provide an account of the goals, methodology,
and subject matter of $X$.
\end{quote}
Before attempting to give an account of the goals and methodology of a theory of fuzzy computation,
it is necessary to explain what is the subject matter of fuzzy computation. And before explaining what is
fuzzy computation, it is absolutely necessary to explain why fuzzy computation does matter. In
different words, why should we consider fuzzy computation as something important? Naturally,
when it is made clear why fuzzy computing matters, then it would not be {\em easy} for anyone to 
argue that the theory of fuzzy computation should not be part of a general theory of computation. By
admitting the omnipresence of vagueness, then it does make sense to argue that a theory of fuzzy 
computation {\em is} a general theory and, consequently, the ordinary theory of computation is 
a special case of this theory.

Unfortunately, the term {\em fuzzy computation} is considered to be a collective term that describes
fuzzy arithmetic, fuzzy databases, fuzzy web searches, etc. However, these might be considered as
applications of fuzzy computation, but they do not form the body of a theory of fuzzy computation. 
For example, a fuzzy database operates in a non-vague environment and is supposed to handle vague data. 
Clearly, this is quite  useful, but it is far from being a vague computation. In fact, to say that
a fuzzy database is vague computing is is like saying that a simulator of a quantum computer is can 
achieve exactly what a real quantum computer can. A theory of fuzzy computation
should propose methods to compute in a vague environment (e.g., conceptual computing devices that 
operate in an environment, where, for instance, one cannot precisely measure the position of particles, 
and their operations are vague) that ``realize'' vagueness using fuzzy set theory.  A theory of fuzzy 
computation could be also used to examine whether there are limits to what can be computed. Since Turing 
machines are the archetypal conceptual computing device, they have been used as a basis to define fuzzy 
conceptual computing devices.

Any new theory should not ignore predictions and results delivered by older theories. For example, 
consider the general theory of relativity. This theory predicted things that were predicted by Newton's 
theory of gravity, nevertheless, it was able to make predictions about things where Newton's theory 
failed. Thus, a theory of fuzzy computation should enrich current theories by incorporating vagueness 
into them.

\paragraph{Plan of the paper} Before providing an account of the goals and methodology of fuzzy
computation, it is necessary to explain what is vagueness, in general, and fuzziness, in particular,
and then to give an overview of fuzzy Turing machines.
\section{What is Vagueness?}
In the English language the word {\em fuzzy} is a synonym of the word {\em vague}. Typically, the
term {\em vague} is used to denote something uncertain, imprecise or ambiguous. Nevertheless,
it is widely accepted that a term is vague to the extent that it has borderline cases, that is, 
cases in which it seems impossible either to apply or not to apply this term. The Sorites Paradox, 
which was introduced by Eubulides of Miletus, is a typical example of an argument that shows what it is 
meant by borderline cases. Also, the paradox is one of the so called {\em little-by-little 
arguments}.  The term sorites derives from the Greek word for heap. The paradox is about the number of 
grains of wheat that makes a heap. All agree that a single grain of wheat does not comprise a heap. 
The same applies for two grains of wheat as they do not comprise a heap, etc. However, there is a point 
where the number of grains becomes large enough to be called a heap, but there is no general agreement 
as to where this occurs.

Bertrand Russell~\cite{russell1923} was perhaps the
first thinker who gave a definition of vagueness:
\begin{quote}
{\em Per contra}, a representation is {\em vague} when the relation of the
representing system to the represented system is not one-one, but one-many.
\end{quote}
Based on this definition, one can conclude that a photograph that is so smudged, it might equally 
represent Brown or Jones or Robinson, is actually a vague photograph. Building on Russell's ideas, 
Max Black~\cite{black1937} had argued that most scientific theories, and of course any theory of computation
should not excluded,  are ``ostensibly expressed in terms of objects never encountered in
experience.'' Thus, one could argue that the Turing machine is an idealization of some real-world system. 
Consequently, the important question is whether it does correspond to anything real. I plan not to answer
this question here so do not expect an answer in what follows. Black~\cite{black1937} proposed as a definition of vagueness 
the one given by Charles Sanders Peirce:
\begin{quote}
A proposition is vague when there are possible states of things concerning which it is 
intrinsically uncertain whether, had they been contemplated by the speaker, he would have 
regarded them as excluded or allowed by the proposition. By intrinsically uncertain we 
mean not uncertain in consequence of any ignorance of the interpreter, but because the 
speaker's habits of language were indeterminate. 
\end{quote}
According to this definition, vagueness is a linguistic phenomenon, but also real life objects exhibit some sort
vagueness. For instance, clouds are definitely vague objects as one cannot precisely specify their 
boundaries. In addition, the ``orbit'' of an electron is a cloud with no precise boundaries. However,
I will say more on vague objects on page~\pageref{vague:real}. Readers interested in a general 
discussion of vagueness should consult a book like~\cite{vague:book}. As far it regards the linguistic 
phenomenon, one could argue that although the term dirty is a linguistic term, still when it refers to 
cloths and the effort needed to clean them, the term ceases to be just a linguistic term and describes 
a real-world situation. Furthermore, washing machines that employ these ``linguistic'' terms to clean our 
cloths by making reasonable use of energy, water, and detergent are not some imaginary things, but real 
appliances available to everyone. Of course, one could argue further that there is no vagueness at all 
and we can precisely describe dirt using some sort of scale. Indeed, but then we are talking about degrees of 
dirtiness, something that is chiefly modeled with fuzzy sets. For a more detailed discussion of these and
other arguments against vagueness see~\cite{syropoulos14}.

It is widely accepted that there are (at least) three different expressions of 
vagueness~\cite{sep-vagueness}:
\begin{description}
\item[Many-valued Logics and Fuzzy Logic] Borderline statements are assigned truth-values that 
                  are between absolute truth and absolute falsehood (see~\cite{smith2008} for a book-length
                  discussion of this idea).
\item[Supervaluationism] The idea that borderline statements lack a truth value.
\item[Contextualism] The truth value of a proposition depends on its context (i.e., 
      a person may be tall relative to American men but short relative to NBA players).
\end{description}
There is a fourth, more recent, expression of vagueness that is based on the use of paraconsistent logics~\cite{vague:book,hyde2008}.
In what follows, I will discuss only models of vague computation where vagueness is described using fuzzy set theory. Therefore, it is
necessary to outline the corresponding ideas. 

\section{Fuzzy Set Theory in a Nutshell}
Fuzzy set theory was introduced by Lotfi Askar Zadeh~\cite{zadeh1965} as an extension of
ordinary set theory. Zadeh defined fuzzy sets by generalizing the membership relationship. In 
particular, given a universe $X$, he defined a fuzzy subset of $X$ to be an object that is
characterized by a 
function $A:X\rightarrow[0,1]$. Any value $A(x)$ specifies the degree to which an element 
$x$ belongs to $A$. Let me now present the basic operations between 
fuzzy subsets.

Assume that $A,B:X\rightarrow[0,1]$ are two fuzzy subsets of $X$. Then, their union and
their intersection are defined as follows:
\begin{align*} 
(A\cup B)(x)&=\max\{A(x),B(X)\}\\ 
\intertext{and}
(A\cap B)(x)&=\min\{A(x),B(X)\}.
\end{align*}
Also, if $\bar{A}$ is 
the complement of the fuzzy subset $A$, then $\bar{A}(x)=1-A(x)$. More generally, it is quite 
possible to use functions other than $\min$ and $\max$ to define the intersection and the union 
of fuzzy subsets. These functions are known in the literature as {\em t-norms} and {\em t-conorms},
respectively.
\begin{definition}
A t-norm is a binary operation $\ast:[0,1]\times[0,1]\rightarrow[0,1]$ that satisfies at least 
the following conditions for all $a,b,c\in[0,1]$:
\begin{description}
\item[Boundary condition] $a\ast 1=a$ and $a\ast 0=0$.
\item[Monotonicity] $b\le c$ implies $a\ast b \le a\ast c$.
\item[Commutativity] $a\ast b=b\ast a$.
\item[Associativity] $a\ast(b\ast c)=(a\ast b)\ast c$.
\end{description}
\end{definition}
\begin{definition}
A t-conorm is a binary operation $\star:[0,1]\times[0,1]\rightarrow[0,1]$ that satisfies at least 
the following conditions for all $a,b,c\in[0,1]$:
\begin{description}
\item[Boundary condition] $a\star 0=a$ and $a\star 1=1$.
\item[Monotonicity] $b\le c$ implies $a\star b \le a\star c$.
\item[Commutativity] $a\star b=b\star a$.
\item[Associativity] $a\star(b\star c)=(a\star b)\star c$.
\end{description}
\end{definition}
For more information on t-norms and t-conorms see~\cite{klir95} or any other textbook on
fuzzy set theory.

\section{The Subject Matter}
Essentially, the classical theory of computation started with the publication of Alan Turing's
{\em On Computable Numbers, with an application to the Entscheidungsproblem}~\cite{turing36}.
Turing's paper introduced a conceptual computing device, that now bears his name, which was
devised in order to solve the Entscheidungsproblem, that is, a problem that was put forth by
David Hilbert in 1928. Roughly speaking, the Entscheidungsproblem asks if it is possible
to find a method that will take as input a description of a formal language and a 
statement in the language and produce as output either {\em True} or {\em False} 
according to whether the statement is true or false. 

The Turing machine is a very simple conceptual computing device that consists of an infinite 
tape, which is divided into writable cells, a scanning head that can read the contents of
a cell or print something to a cell, and the so-called {\em controlling device}, which is
a lookup table that controls the behavior of the machine. Initially, one writes the
input data into the cells of the tape, and then sets the machine into motion. The machine 
delivers a result if it stops after some finite amount of time. This simple machine is
surprisingly powerful and it can be used to compute many functions and numbers. Not so
surprisingly, the Turing machine is considered as the cornerstone of the (classical)
theory of computation. 

When introducing a new model of computation, it is almost customary to introduced some
sort of Turing machine that accommodates the central idea behind the new model of computation.
For example, in order to introduce probabilistic computing, Eugene Santos~\cite{santos1971}
introduced a probabilistic Turing machine. The same principle applies to quantum
computing~\cite{hirvensalo2004}. However, not all Turing machine counterparts
are very helpful. For instance, in quantum computing it is common to use quantum circuits in various 
theoretical studies because they are more expressive in describing quantum algorithms. 

The first steps towards the definition of a fuzzy Turing machine have been made by the
inventor of fuzzy set theory Zadeh~\cite{zadeh1968}. Unfortunately, Zadeh did not
provide a definition of a fuzzy Turing machine or of a fuzzy algorithm. One could say
that he actually described what a fuzzy algorithm might be and, in a sense, speculated
about the properties of a fuzzy Turing machine. However, his work prompted other researchers
to investigate the notions of fuzzy Turing machines and fuzzy algorithms (see~\cite{syropoulos14}
for a comprehensive account of all these formulations). Although, fuzzy Turing machines are
not the only model of fuzzy computation and, to some extend, not the most natural one, still it has
been studied thoroughly and thus it makes sense to give the definition of a fuzzy Turing machine.

The definition that follows was proposed Ji{\v{r}\'\i} Wiedermann~\cite{wiedermann2004} and
I consider it the most complete and general definition of a fuzzy Turing machine:
\begin{definition} 
A nondeterministic fuzzy Turing machine with a unidirectional tape is a nonuple 
\begin{displaymath}
\mathcal{F}=(Q,T,I,\Delta,\openbox,q_0,q_f,\mu,\ast),
\end{displaymath}
 where:
\begin{itemize}
\item $Q$ is a finite set of states;
\item $T$  is a finite set of tape symbols;
\item $I$ is a set of input symbols, where $I\subseteq T$;
\item $\Delta$ is a transition relation and it is a subset of $Q\times T\times Q\times T
\times \{L,N,R\}$. Each action that the machines takes is associated with an element 
$\delta\in\Delta$. In particular, for $\delta=(q_i,t_i,q_{i+1},t_{i+1},d)$ this means that 
when the machine is in state $q_i$ and the current symbol that has been read is $t_i$, then
the machine will enter state $q_{i+1}$, the symbol $t_{i+1}$ will be printed on the
current cell and the scanning head will move according to the value of $d$, that is, if
$d$ is $L$, $N$, or $R$, then the head will move one cell to the left, will not move, or it
will move one cell to the right, respectively.
\item $\openbox\in T\setminus I$ is the blank symbol;
\item $q_0$ and $q_f$ are the initial and the final state, respectively;
\item $\mu:\Delta\rightarrow[0,1]$ is a fuzzy relation on $\Delta$; and
\item $\ast$ is a t-norm. 
\end{itemize}
\end{definition}

In order to fully understand how this conceptual machine works, it is necessary to present
a few additional notions.
\begin{definition}
When $\mu$ is partial function from $Q\times T$ to $Q\times T\times \{L,N,R\}$ and $T$ is a
fuzzy subset of $Q$, then the resulting machine is called a
deterministic fuzzy Turing machine.
\end{definition}

A configuration gives the position of the scanning head, of what is printed on the tape, and
the current state of the machine. If $S_i$ and $S_{i+1}$ are two configurations, then 
$S_{i}\vdash^{\alpha} S_{i+1}$ means that $S_{i+1}$ is reachable in one step from
$S_{i}$ with a plausibility degree that is equal to $\alpha$ if and only if there is a
$\delta\in \Delta$ such that $\mu(\delta)=\alpha$ and by which the machine goes from $S_i$ 
to $S_{i+1}$. When a machine starts with input some string $w$, the characters of the string 
are printed on the tape starting from the leftmost cell; the scanning head is placed atop the
leftmost cell, and the machine enters state $q_0$. If
\begin{displaymath}
S_{0}\vdash^{\alpha_0} S_{1}\vdash^{\alpha_1} S_{2}\vdash^{\alpha_2}\cdots\vdash^{\alpha_{n-1}} S_{n},
\end{displaymath} 
then $S_n$ is reachable from $S_0$ in $n$ steps. Assume that $S_n$ is reachable from $S_0$ in
$n$ steps, then the plausibility degree of this 
{\em computational path} is
\begin{displaymath}
D\bigl( (S_{0},S_{1},\ldots,S_{n})\bigr)=\left\{ \begin{array}{ll}
                                                    1, & n=0\\
                     D\bigl( (S_{0},S_{1},\ldots,S_{n-1})\bigr)\ast\alpha_{n-1}, & n>0
                                                 \end{array}\right.
\end{displaymath}
Obviously, the value that is computed with this formula depends on the specific path that is
chosen. Since the machine is nondeterministic, it is quite possible that some configuration
$S_n$ can be reached via different computational paths. Therefore, when a machine starts from
$S_0$ and finishes at $S_n$ in $n$ steps, the plausibility degree of this computational path,
which is called a {\em computation}, should be equal to the maximum of all possible computation paths:
\begin{displaymath}
d(S_n)=\max\Bigl[ D\bigl( (S_{0},S_{1},\ldots,S_{n})\bigr) \Bigr].
\end{displaymath} 
In different words, the plausibility degree of the computation is equal to the plausibility
degree of the computational path that is most likely to happen.

Assume that a machine starts from configuration $S_0$ with input the string $w$. Then, a
computational path $S_0,S_1,\ldots,S_m$ is an accepting path of configurations if the
state of $S_m$ is $q_f$. In addition, the string $w$ is accepted with degree equal to
$d(S_m)$.
\begin{definition}
Assume that $\mathcal{F}$ is a fuzzy nondeterministic
Turing machine. Then, an input string $w$ is accepted with plausibility degree $e(w)$ by 
$\mathcal{F}$ if and only if:
\begin{itemize}
\item there is an accepting configuration from the initial configuration $S_0$ on input $w$;
\item $e(w)=\max_{S}\bigl\{d(S)\mathrel{|}\text{$S$ is an accepting 
configuration reachable from $S_0$}\bigr\}$.
\end{itemize}
\end{definition}
Also,
\begin{definition}\label{fuzzy:accept:crit}
The fuzzy language accepted by some machine $\mathcal{F}$ is the fuzzy set that is defined as
follows:
\begin{displaymath}
L(\mathcal{F})=\Bigl\{ \bigl(w,e(w)\bigr)\mathbin{\Bigm|} \text{$w$ is accepted by $\mathcal{F}$ 
with plausibility degree $e(w)$}\Bigr\}.
\end{displaymath}
\end{definition}

It is quite possible that a Turing machine might not write on its tape the symbol 1 but something that
looks like it. Then, when reading this symbol the machine might have trouble deciding whether it
is an 1 or not. Obviously, this a very intresting idea but it might happen because the read-write
head is defective. If this happens randomly, then it should be necessary to include this ``feature''
in our analysis. However, this is an open problem at the moment. 

Another model of fuzzy computation, more close to the idea of vagueness, are fuzzy P~systems 
(see~\cite{syropoulos14} for an up-to-date and thorough presentation  of fuzzy P~systems and other
similar fuzzy models of computation). P~system have been introduced by Gheorghe P{\u{a}}un~\cite{paun2002}.
Roughly, a P~system is a model of computation that is based on the functionality of the cell.
To the best of my knowledge, any cell has a membrane that surrounds it, separates 
its interior from its environment, regulates what goes in and out, etc. Inside the membrane, 
the cytoplasm takes up most of the cell volume. Various organelles (i.e., specialized 
subunits within a cell that have a specific function) are ``floating'' inside the cytoplasm. 
Just like a cell, a P~system is enveloped in a porous membrane that allows objects to move 
in and out. Inside a P~system there  is an indefinite number of nested compartments,
that is, compartments that may contain other compartments, etc., each of them enveloped by 
a porous membrane. Also, there is a designated compartment called the {\em output
compartment}. In addition, each compartment, may contain ``solid,'' possibly repeated objects,
that is, a multiset of objects, while it is associated with a set of multiset rewriting rules. 
The system operates in discrete time and these rules specify what changes can possible happen
inside a compartment at each tick of the clock. In general, compartments cannot be deleted 
while objects may be multiplied, deleted, or introduced in a compartment. Computation stops when 
no rule is applicable and the result of the computation equals the number of objects that have 
been accumulated in the output compartment. Now, a fuzzy P~system is one where the multisets of
objects are replaced by fuzzy multisets:
\begin{definition}
Assume that $M:X\rightarrow\mathbb{N}$ characterizes a multiset $M$.
Then, a fuzzy multi-subset of $M$ is a structure $A$ that is
characterized by a function $A:X\rightarrow \mathbb{N}\times [0,1]$
such that if $M(x)=n$, then $A(x)=(n,i)$. In addition,
the expression $A(x)=(n,i)$ denotes that the degree to which each of the
$n$ copies of $x$ belong to $A$ is $i$.
\end{definition}
The cardinality of such a set is given by the following formula:
\begin{displaymath}
\card A=\sum_{a\in A} A_{m}(a) A_{\mu}(a),
\end{displaymath}
where $A_{m}(a)=n$ and $A_{\mu}(a)=i$.
\begin{corollary}
The result of a computation delivered by a fuzzy P~system is a positive real number.
\end{corollary}
\section{Goals and Methodology}
Philosophical questions regarding the distinction between hardware and software and other similar questions
are essentially the same when it comes to a philosophy of fuzzy computation. However, the important question
here is where vagueness comes into the picture when one deals with real computers? In different words, is the
hardware we use today non-exact or non-well-defined so that it is justified to see vagueness come into play? 

It should not surprise anyone the fact that modern computers consume electricity that might be fluctuating. 
Typically, machines can cope with some small fluctuations of electricity because, among others, they have been 
designed to operate within a specific range of voltage. These fluctuations of electricity and other similar
phenomena (e.g., noise in communication) are some sort of vague phenomena and, therefore, a raison d'\^{e}tre for
vagueness. However,  we intensionally choose to ignore this sort of vagueness and assume that our systems
are exact when they are roughly exact. Of course this happens because those who built the first computers
where not interested in vagueness and imprecision but rather in exactness and precision. Thus, one can
safely conclude that there is some sort of vagueness in hardware. The next big question is how can we make use 
of this vagueness? Or, going one step further, what would make a computer vague?

Before trying to answer this question, it is necessary to see whether it actually makes sense to expect to see
the construction of a vague computer. Martin Davis~\cite{davis1988} speculated that John von Neumann was aware
of Turing's work and that he has used it in the construction of ENIAC. Of course this is an arbitrary conclusion,
as is based on a speculation. After all, Konrad Zuse's Z3 computer~\cite{z3} was not build on ideas borrowed from 
Turing although he was a contemporary of Turing too. As far it regards modern computers, it is known that 
{\em interaction} (see~\cite{syropoulos08}) is a notion ``alien'' to the Turing model of computation. This implies 
that the construction of a vague computing machine should not be necessarily based on the properties of a fuzzy 
Turing machine. However, this remark is quite important because of a crucial theoretical result.

Although some authors stubbornly insist that modern computers are actually universal Turing machines~\cite{petzold08}, 
Selim  G.~Akl~\cite{akl2006,akl2007} denies the existence of universal machines by arguing that only a machine 
that can perform an infinite number of  operations per step can be classified as a universal machine. Even if
one dismisses Akl's position, still Yong Ming Li~\cite{li2006} has proved that there is no universal fuzzy Turing
machine. Interestingly, Li~\cite{li2008b} has also concluded that there is a restricted form 
of a universal fuzzy Turing machine. However, it seems that this conclusion is somehow a biased one, mainly
because people cannot accept that there is no universal fuzzy Turing machine. Thus if we accept that every
real computing device is a realization of some universal conceptual computing device, one could not build
a real machine based on the properties of the fuzzy Turing machine. This means that one should seek inspiration
to other models of vague computation, provided there are universal versions of these models. Alternatively,
one could just ignore this ``requirement'' and try nevertheless to build a machine. Surprisingly, based on remarks 
about the ``fuzziness'' of quantum mechanics (e.g., see~\cite{granik2001,pykacz2011,pykacz2001}), one could argue 
that quantum computers are  actually vague computers, provided one replaces probability theory with with possibility
theory or any other theory derived from fuzzy set theory. However, even if this would be proved to be a fallacy, 
which is very unlikely, there is another way out---Vague computing devices should make use of vagueness as is manifested 
in the molecular, atomic or subatomic level. Since it is quite possible that some (?) readers may find almost absurd the 
idea that vagueness exists at the quantum level, I would like to present an example that hopefully sheds light on this idea.
The following example, which was originally presented by E. J.~Lowe~\cite{lowe1994}, shows that vagueness exists in the
subatomic level:
\begin{quote}\label{vague:real}  
Suppose (to keep matters simple) that in an ionization chamber a
free electron $a$ is captured by a certain atom to form a negative ion which,
a short time later, reverts to a neutral state by releasing an electron $b$. 
As I understand it, according to currently accepted quantum-mechanical    
principles there may simply be no objective fact of the matter as to whether 
or not $a$ is identical with $b$. It should be emphasized that what is 
being proposed here is not merely that we may well have no way of telling whether 
or not $a$ and $b$ are identical,which would imply only an epistemic indeterminacy.
It is well known that the sort of indeterminacy presupposed by orthodox
interpretations of quantum theory is more than merely epistemic---it is
ontic. The key feature of the example is that in such an interaction electron
$a$ and other electrons in the outer shell of the relevant atom enter an 
`entangled' or `superposed' state in which the number of electrons present 
is determinate but the identity of any one of them with $a$ is not, thus 
rendering likewise indeterminate the  identity of $a$ with the released electron
$b$.
\end{quote}
The idea behind this example is that ``identity statements represented by `$a=b$' are `ontically' 
indeterminate in the quantum mechanical context''.\footnote{Steven French and Décio Krause.
\newblock {\em Quantum Vagueness}.
\newblock Erkenntnis, 59,  pp.~97--124, 2003.} In different words, in the 
quantum mechanical context $a$ is equal to $b$ to some degree, which is one of the fundamental 
ideas behind fuzzy set theory. Based on this observation, it should be clear that a vague computer should make
use of vagueness as is manifested in nature. 

A vague computer should be able to run programs. Clearly, such a computing device should be able to run conventional
programs, since, as expected, all vague models of computation would be far more general than their {\em crisp} 
counterparts. Forget for a moment the connection between vague and quantum computation and assume that
there are no vague computers  available at this moment. Then, does it make sense to talk about vague
programs and vague programming languages today? The answer is emphatically {\em yes} and here is the rationale. Today,
quantum computers are not widely available yet, still there are quantum programming languages, 
like Quipper~\cite{quipper2013}, which can be used to create quantum programs. Such programming
languages are implemented using conventional methodologies and techniques (e.g., Quipper is written in Haskell)
and programs expressed in these languages run on conventional hardware. Similarly, one can design
and implement a vague programming language using a conventional programming language so that
vague programs run on conventional hardware. However, one should note that since quantum computers are more 
powerful than conventional machines (e.g., they can compute true random numbers~\cite{pironio2010} whereas
von Neumann machines can compute only pseudo-random number), it is of course impossible to use this extra 
power when working with such quantum programming languages. And of course this applies to vague programming languages 
too, provided that vague machines have similar capabilities.

Zadeh~\cite{zadeh1968} has speculated about the form of commands in a vague programming language. Thus, according
to Zadeh, a typical command of such a language would be ``set $y$ approximately equal to 10 if $x$ is approximately equal 
to 5.'' Ever since a number of vague programming languages have been designed and implemented. RASP~\cite{fuzzyproglang1} 
was an extension of BASIC that provided basic operations for fuzzy sets but it did not include commands similar to the ones
suggested by Zadeh's. The FLISP~\cite{fuzzyproglang2} programming language, an extension of LISP, provided facilities to input 
and process fuzzy data. For example, in order to enter the fuzzy set 
\begin{displaymath}
Q = 0.3/2 + 0.9/3 + 1/4 + 0.8/5 + 0.5/6
\end{displaymath}
where $f/d$ means that $d\mathrel{\in_{f}}Q$, one had to enter the following commands:
\begin{verbatim}
     (SETQ U '( 0 1 2 3 4 5 6 7 8 9))
     (FSETQ Q ((U) (FSET ((2@0.3)(6@0.5)(5@0.8)(3@-0.9)(4@1)))))
\end{verbatim}
HALO~\cite{fuzzyproglang3} was a LISP-like language that employed many Pascal-like and C-like structures.
In addition, the logical operations as well as some other operations are fuzzy. The assignment statement of the
languages L and XL~\cite{fuzzyproglang4} allowed users to assign fuzzy numbers to variables. In addition,
XL included a fuzzy repetition construct. Fril++~\cite{fuzzyproglang5} is an object-oriented language 
where an object can be an instance of a class to some degree. This is a particularly interesting idea and
implies that two instances of some class can be equal to some degree. And it would be quite interesting
to see how one could implement this idea in a way similar to Java's \texttt{equals()} method. Fuzzy Arden 
Syntax is programming language that has been designed ``to provide an easy means of 
processing vague or uncertain data, which frequently appears in medicine''~\cite{fuzzyproglang6}. 
One can define fuzzy sets very easily:
\begin{verbatim}
     U:= fuzzy set (2,0.3), (6,0.5), (5,0.8), (3,-0.9), (4,1);
\end{verbatim}
The language allows commands that are reminiscent of Zadeh's ``commands'':
\begin{verbatim}                   
     TempatureList:= read {temperature} where
       it occurred within the past 24 hours
       fuzzified by 4 hours;
\end{verbatim}
And of course, there is a fuzzy $\lambda$-calculus~\cite{fuzzyproglang7} where each term is associated with
a degree and the b-reduction is redefined. I suppose the authors meant $\beta$-reduction, but that is just 
an educated guess\ldots The notion of ``approximately equal'' can be introduced in a language by means of an 
extended assignment operator (e.g., \verb|~=|) and an extended equality operator (e.g., \verb|~==|). Thus, a command like
\begin{center}
\verb|x ~= y|
\end{center}
would mean that \texttt{x} is actually assigned a compound value like $\mbox{\texttt{y}}\pm\delta y$, where the
$\delta y$ should be implementation dependent or the user should be able to configure the compiler accordingly. 
In a sense, \texttt{x} would an interval and not just a point. 

The languages that were briefly reviewed can be roughly divided into two categories: those that allow the use
of fuzzy sets and those that implement some basic principle of fuzzy set theory (e.g., similarity of objects).
The languages that implement some sort of similarity with degree are closer to the spirit of vagueness. In 
general, fuzzy programming languages should provide facilities for the definition and manipulation of fuzzy
sets. In addition, they should provide control structures that can handle fuzzy logical expression. Also,
it is necessary to provide facilities to express similarities between structures.

In the previous section, I stated that Turing devised his (automatic) machine in order to solve the Entscheidungsproblem.
This is a bit inaccurate. The truth is that Turing, devised his machine and then he devised the universal Turing
machine with which he gave his answer to Entscheidungsproblem. In modern computer parlance,
Turing essentially proved that it is not possible to tell whether a program that is not {\em responding}
has entered a vicious circle or not. This problem is known as the {\em halting problem}. By showing that the
halting problem is as hard as the Entscheidungsproblem, he proved that the Entscheidungsproblem is unsolvable, 
or better it is {\em Turing} unsolvable. Turing went one step further and made a bold statement---if a number or a 
function is computable, then it must be computable by a Turing machine. This statement is now known as 
the {\em Church-Turing thesis}.

Clearly fuzzy Turing machines form an extension of the classical archetypal conceptual computing device.
Therefore, they should compute as many numbers and/or functions as their classical counterpart. The 
important question is whether these machines are more powerful than their classical counterparts. Ji{\v{r}\'\i} 
Wiedermann~\cite{wiedermann2004} has shown that fuzzy Turing machines are more powerful than 
ordinary Turing machines. In different words, fuzzy Turing machines are {\em hypercomputers}~\cite{syropoulos08}. 
Despite this {\em fact}\footnote{See~\cite{syropoulos14} for a discussion of various attacks to this result.}
it is not clear at all what are the computational limits of these machines. After all, for each conceptual and logically
consistent computing device there is a limit to what it can achieve. For example, Toby Ord and Tien 
D.~Kieu~\cite{ord05} have argued that every logically consistent computing device cannot solve its
halting problem. Since there are no universal fuzzy Turing machines, it makes no sense to talk about their
halting problem, nevertheless, it does make sense to try to find their computational limits.

Mark Changizi~\cite{changizi2003} assumes the validity of three theses or hypotheses: the Church-Turing thesis,
the Programs-in-Head hypothesis, and the Any-Algorithm hypothesis.  These hypotheses state:
\begin{quote}
\textbf{Programs-in-Head Hypothesis} For most natural language predicates $P$ and
their natural language negation `$\neg P$', their interpretations are determined
by you using programs in the head.
\end{quote}
\begin{quote}
\textbf{Any-Algorithm Hypothesis} You are free to choose from the set of all algorithms 
when interpreting natural language predicates or their natural language negations.
\end{quote}
Changizi argues that these hypotheses together with the Church-Turing thesis imply the
omnipresence of vagueness in language. Of course, it is not know if the Church-Turing
thesis is valid and hypercomputation implies that it is not. Next, the two hypotheses 
are based on the assumption that mechanism is valid, which is equally problematic.
In~\cite{syropoulos08} I have argued against mechanism so I will not repeat these arguments here.
What is even more problematic is that vagueness is restricted into language and because of
Changizi's arguments any language is mostly vague. However, on a place with no intelligent beings
there is no language and os no vagueness! Ergo, vagueness is not something real. And this is
the reason I have tried to establish that vagueness is a fundamental property of the physical
world and notr some human invention.

\section{Conclusions}
Fuzzy computing is a new branch of (theoretical) computer science that is not fully developed. There are a number of
open questions regarding fuzzy conceptual computing devices and their capabilities. The answers to these
questions greatly depends on one's philosophical prejudices. I have tried to briefly present the field
of fuzzy computation and to discuss these open problems based on my own prejudices. In summary, there is
no universal fuzzy Turing machine and it seems universality has nothing to do with vague computing
devices. However, this should not pose an obstacle in the construction of vague computers, which should be
based on vagueness as it appears at the particle level. Also, it seems that vague computing devices will be more 
powerful than their classical counterparts, but the upper bound of their computational power has not been determined yet.  

\end{document}